\documentclass[twocolumn,showpacs,showkeys,preprintnumbers,amsmath,amssymb,nofootinbib]{revtex4-1}
\usepackage[latin1]{inputenc}         
\usepackage[english]{babel}
\usepackage{bm}                           
\usepackage{graphicx,color}
\usepackage{amsmath}
\usepackage{epsfig}                       
\begin{document}
\hyphenation{con-se-cu-tive}
\def\be{\begin{equation}}
\def\ee{\end{equation}}
\def\lb{\label}

\draft 





\title[JOURNAL OF APPLIED PHYSICS]
{Magnetic Phases and Specific Heat of Ultra-Thin Holmium Films}

\author{L.J. Rodrigues,}
 \affiliation{Departamento de F\'{\i}sica Teórica e Experimental, Universidade Federal do Rio
Grande do Norte, Natal - RN 59600-900, Brazil}

\author{V.D. Mello}
\affiliation{Departamento de F\'{\i}sica, Universidade do Estado do Rio Grande do Norte, Mossor\'o - RN 59625-620, Brazil}

\author{D.H.A.L. Anselmo}
\affiliation{Departamento de F\'{\i}sica Teórica e Experimental, Universidade Federal do Rio Grande do Norte, Natal - RN 59072-970, Brazil}

\author{M.S. Vasconcelos}\footnote{Phone: +55 84 3342 2355; Fax: +55 84 3215 3791} \email{manoelvasconcelos@yahoo.com.br; mvasconcelos@ect.ufrn.br} \affiliation{Escola de
Ciência e Tecnologia, Universidade Federal do Rio Grande do Norte,
59072-970, Natal- RN, Brazil}

\date{\today}

\begin{abstract}

 We report model calculations of the magnetic phases of  very thin Ho films in the temperature interval between 20K and 132K, and show
that slab size, surface effects and magnetic field due to spin ordering may impact significantly the magnetic phase diagram. There is a relevant reduction of the external field strength required to saturate the magnetization and for ultra-thin films the helical state does not form. We explore the heat capacity and the susceptibility as auxiliary tools to discuss the nature of the phase transitions.

\end{abstract}



\maketitle 


\section{Introduction}

Nowadays, rare earths are vital to some of the world's fastest
growing markets: clean energy and high technology (for a review, see
\cite{Ronning}, and references therein). They are found in most
everyday applications because of their unique chemical and physical
properties. New applications have arisen consistently over the past
50 years, including important environmental innovations such as
catalytic converters and the development of permanent magnets which
have enabled greater efficiency, miniaturization, durability and
speed in electric and electronic components. Generally they are
classified into one of two categories, namely, light rare earths and
heavy rare earths, with varying levels of uses and demand on the
industry technologies. For example, the lanthanum is a light rare
earth responsible for making rechargeable lanthanum nickel metal in
hydride batteries \cite{Binnemans} used in electric and hybrid
vehicles, laptops, cameras, etc. Also it is used to improve visual
clarity in camera lenses, telescopes and binoculars. Also, in optical
fibres it increase significantly the transmission rates. Another heavy rare earth is
the Dysprosium \cite{Brown}, most commonly used in the manufacture
of neodymium-iron-boron high strength permanent magnets. Dysprosium
is used in radiation badges to detect and monitor radiation
exposure. However, the applications in spin wave magnetic devices is
yet scarce.

One of most recent purposes is the use of these materials in solid-state rare-earth-ion-doped
systems, which justifies their status as very strong candidate to a long-lived quantum memory system
\cite{timoney}. Also, very recently, Lovric et al. \cite{marko}, have reported a high fidelity
optical memory, made of a rare earth doped crystal, in which dynamical decoupling is used to
extend the storage time. It is notable, until we know, the lack of studies about the magnetic phases in
rare earth. With this work we intend to fill some part of this lack with the study of the magnetic
phases of rare-earth thin films, aiming further applications.

Magnetic ordering in the heavy rare-earth lanthanides \cite{Koehler} is
mediated by the RKKY interaction in which the polarization of conduction
electron yields an indirect exchange between localized 4f
moments on neighboring lattice sites. The interplay between this long-range interaction and anisotropic and magneto-crystalline effects results in complex magnetic ordering and the possibility of a variety of magnetic structures. As stated before, understanding the magnetism of thin  films is of general importance in a technological context, and rare-earth metals are used in a wide, range of applications, and have attained global economics importance \cite{USEDA,reuters}. Among the rare-earths, Holmium in particular displays unusual behavior which, coupled with the large temperature stability of the intermediate helimagnetic phase when compared to Dy and Tb, has led to it being treated as a model system in a number of recent experimental \cite{Ott} and theoretical \cite{cinti} studies.
In the bulk, Holmium metal orders magnetically below $T_{N}$=132K (N\'eel temperature) and the moments in consecutive neighboring basal planes
rotate to give a long-period helical phase. At temperatures below $T_{C}$=20K (Curie temperature) it arranges as a cone with angle of $80.5^{\circ} \leq \theta$$\leq$ $90^{\circ}$, giving rise to a conical phase. When a magnetic field is applied in the basal plane other magnetic structures are observed including a basal plane ferromagnetic phase, fan phase and helifan phase \cite{Jensen,Jensen1}. In thin films when the thickness is comparable with the periodicity of the ordered structure, is expect that even the magnetic
arrangement itself can be strongly modified. For ultrathin films, however, as the number of spin layers is reduced by decreasing the film thickness, additional structures are established including a fan structure and a spin-slip configuration, highlighting the importance of surface and interfacial effects \cite{Jensen2}. This finite-size effect is caused by the reduced number of atoms in the direction perpendicular to the film plane that leads to a decrease of the  total magnetic exchange energy.
Rare-earth helimagnets such as Ho, Dy, and Tb (for a review see \cite{Jensen2}) represent the best candidates to put into evidence such finite-size effects.

This paper is organized as follows. In Sec. II we provide our theoretical framework for the Hamiltonian, which takes into account the hexagonal anisotropy. A discussion about our model and comparison with other works are done in Sec. II. Finally, we summarize our main conclusions in Sec. IV.

\section{Theoretical Model}

We investigate a c-axis thin film, consisting of a stacking of
atomic layers with equivalent spins, infinitely extended in the x-y
directions. The spins in each monolayer are exchange coupled with
the spins in the first and second neighbor monolayers. The
anisotropy is uniform throughout the film and the near surface spins
have reduced exchange energy. The magnetic Hamiltonian is given by:

\begin{eqnarray}\label{eq1}
\textit{H}=J_{1}(g-1)^{2}\sum_{n=1}^{N-1}\vec{J}(n)\cdot\vec{J}(n+1)+\nonumber \\
J_{2}(g-1)^{2}\sum_{n=1}^{N-2}\vec{J}(n)\cdot\vec{J}(n+2)+\\
\sum_{n=1}^{N}\left[K_{6}^{6}\cos(6\varphi_n)-g\mu_B\vec{J}(n)\cdot\vec{H}
\nonumber\right]
\end{eqnarray}

In Eq. (\ref{eq1}),  $J_{1}$ and $J_{2}$ describe the exchange
interaction between the nearest and next nearest monolayers,
$\vec{J}$(n) denotes the total angular momentum per atom in the
$n-th$ monolayer. $K_{6}^{6}$ describes the hexagonal anisotropy
and the last term is the Zeeman Energy, where the external
field, $\vec{H}$, is applied in one easy direction in the hexagonal
plane, making an angle of $30^\circ$ with $x$ axis. Also, $K_{6}^{6}(T)$ is adjusted so as to reproduce the
temperature dependence \cite{Coqblin} of the hexagonal anisotropy
energy. We use the Ho bulk energy parameters \cite{Legvold}, where J=8, $J_{1}=47k_{B}$,
$J_{2}=-J_{1}/4\cos\phi(T)$, where $\phi(T)$ is the temperature
dependent helix turn angle \cite{Andrew}. Also, $g=5/4$  is the Landé
factor, corresponding to a saturation magnetic moment per atom of
9.5$\mu_{B}$. We use a self-consistent local field model that
incorporates the surface modifications in the exchange field and the
thermal average values $(\langle J(n) \rangle $; $n=1; ...N)$ and the orientation
of the spins in each layer $(\langle \phi_{n} \rangle $; $n = 1;
...N)$ \cite{Carrico,Mello}.

\section{Results and Discussion}

In this paper we will concentrate our analysis mainly on
the phase diagrams $H-T$ obtained for Holmium bulk and for the films composed by n=24, 10 and 7 monolayers.
We present the $H$-$T$ diagram in the temperature interval from 20K to 132K. The magnetic phase transitions
are function of the external magnetic field $H$, temperature $T$,
 and thickness $N$. We use the specific heat (magnetic and lattice contributions),
 and the magnetic susceptibility to identify the nature of the magnetic phase transition.

\begin{figure}[h]
\includegraphics[scale=0.26]{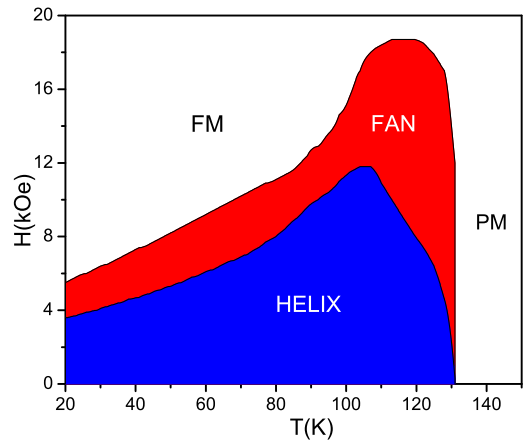}
%
\caption{Phase diagram of Ho bulk.}\label{phase1}
\end{figure}

In the absence of external magnetic field the Ho bulk film arranges in helimagnetic form from 20K
($T_C$, Curie temperature) to 132K ($T_N$, Néel temperature). In Fig.(\ref{phase1}) we show the $H$-$T$
diagram of Ho bulk where, increasing the field in the isothermal process there is a phase transition from helimagnetic phase to fan phase, going to ferromagnetic phase: Helix$\rightarrow$Fan$\rightarrow$FM. For heating process (and isofield process), we can see up to four transitions. For example, at 8kOe: (FM$\rightarrow$Fan$\rightarrow$Helix$\rightarrow$Fan$\rightarrow$PM).

\begin{figure}[h]
\includegraphics[scale=0.3]{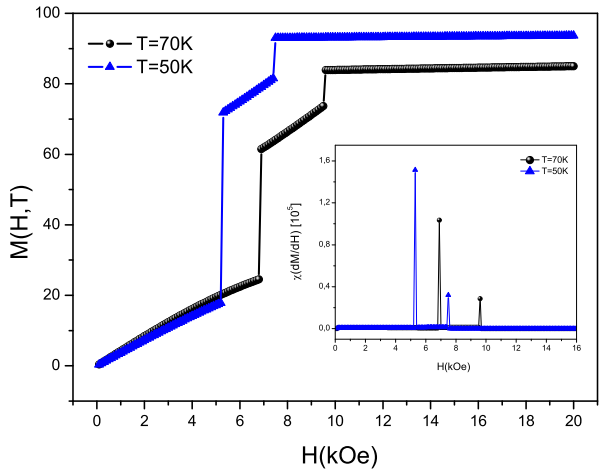}

%
\caption{Isothermal magnetization for T=50K and 70K. {\it Inset: }Magnetic susceptibility for T=50K and 70K.}\label{magsusc}
\end{figure}

In the Fig.(\ref{magsusc}) we show the dependence of isothermal magnetization with applied magnetic field, for both T=50K and 70K. The magnetic susceptibility ($inset$) for selected values of
magnetic field and temperatures is also presented. We can identify in the phase diagram, that for T=50K and T=70K there are two magnetic phase transitions: Helix$\rightarrow$Fan$\rightarrow$FM. For T=50K we can see a peak in the magnetic susceptibility in H=5kOe marking the magnetic phase transition, Helix$\rightarrow$Fan and Fan$\rightarrow$FM in H=7.5kOe

\begin{figure}[h]
%
%
\includegraphics[scale=0.27]{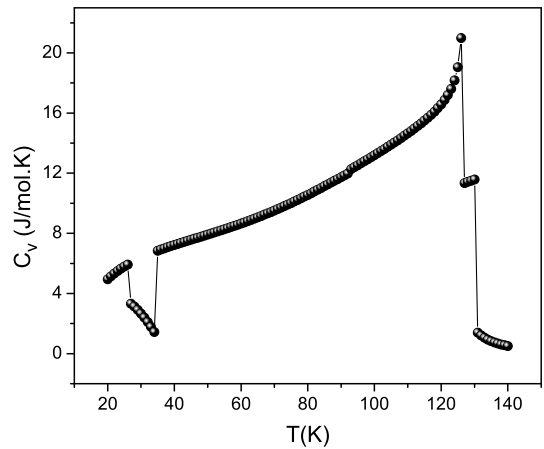}
\caption{Specific Heat of Ho bulk for H=6kOe. Several magnetic phase transitions are observed.}\label{Fig2}
\end{figure}

The specific heat is also investigated, as shown in Fig.(\ref{Fig2}), for selected values of
magnetic field, where both magnetic and
elastic (lattice) contributions are considered. We can see that for H=6kOe, the
magnetic phase transitions are marked by the specific heat. The FM
$\rightarrow$ Fan phase magnetic transition is marked by a deep
peak. The Fan $\rightarrow$ Helix transition exhibit a small plateau
in the specific heat. It is followed by a peak in the
Helix$\rightarrow$ Fan phase magnetic transition and by a drop in
the Fan $\rightarrow$ PM magnetic phase transition. The peak in
specific heat, close to $T_N$, shows the magnetic phase transition,
Fan $\rightarrow$ PM, followed by a drop in the curve due to lattice
contributions. However, the specific heat measurements in weak
magnetic fields do not mark all magnetic phase transitions, because
the field interval which destabilizes the magnetic order is low,
therefore, it is more appropriate to resort to measurements of magnetic susceptibility for fields in that
 interval.

\begin{figure}[h]
%
%
\includegraphics[scale=0.25]{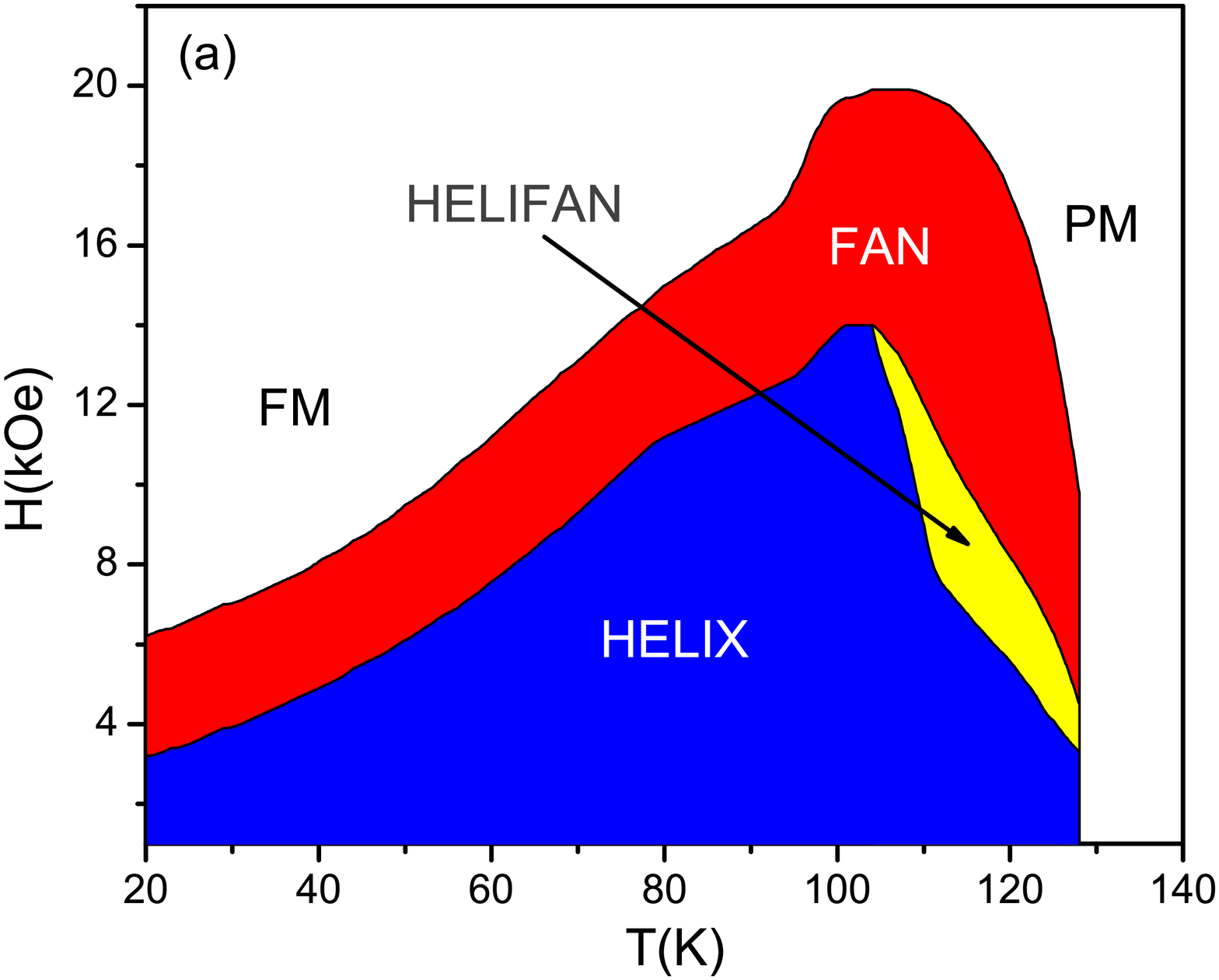}
\begin{center}
\hspace{0.1cm}
\includegraphics[scale=0.32]{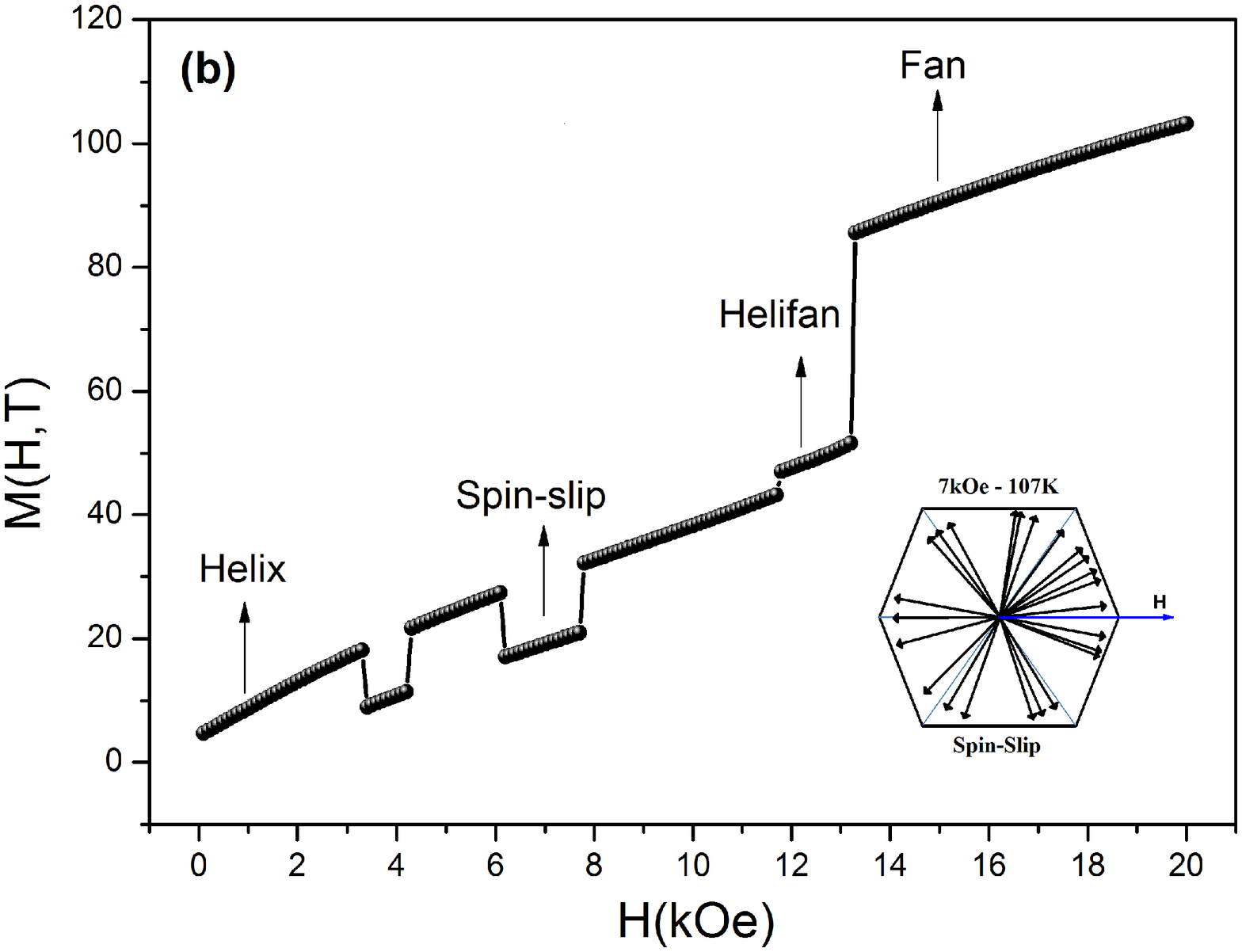}
\caption{(a) Phase diagram of Ho films with 24
monolayers. (b) Isothermal magnetization for $T=107$K. {\it Inset:}
Spin-slip structure for $T=107$K and 7kOe.} \label{spinslipss}
\end{center}
\end{figure}

In Fig.(\ref{spinslipss}-a) we show the $H$-$T$ diagram, describing the magnetic phases for a Ho film consisting of 24
monolayers. In principle, the thickness is enough for two helix
with turning angle of $30^\circ$. But in this case the magnetic phase diagram is
different from the one in bulk. We can see that the presence of surface induces new magnetic phases compared with Ho bulk.
This occurs because the surface spins align more easily with the easy axis of the hexagonal
anisotropy than spins from inner atomic layers, due to the reduced exchange field in the surface region.

The helifan phase is clearly induced by surfaces. There is a threshold
thickness for the stabilization of the helifan magnetic phase. We can see
that there is a short region in the diagram referring to the helifan phase in
the temperature interval from T=103K to 128K and the magnetic field interval from H=3kOe to 13.5kOe.

In Fig.(\ref{spinslipss}-b) we can identify six magnetic phase transitions. The curve shows four peaks in
H=3.1kOe, H=4kOe, H=6kOe and H=7.8kOe for the magnetic phase transitions
Helix$\rightarrow$spin-slip and spin-slip$\rightarrow$Helix.
Also, we can see two peaks at H=11.8kOe and H=13kOe for the magnetic phase transitions
Helix$\rightarrow$Helifan and Helifan$\rightarrow$Fan, respectively. The last phase transition,
Fan$\rightarrow$FM, is marked by a small plateau in H=20kOe (not shown in the figure).
In the inset we illustrate how the magnetic moments arrange in the spin-slip magnetic structure, for T=107K.

\begin{figure}[h]
%
%
\includegraphics[scale=0.30]{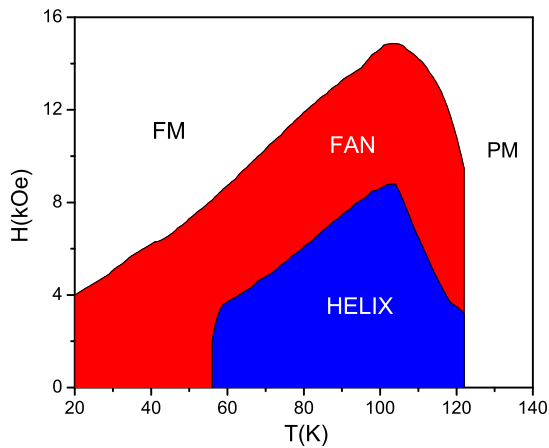}
\caption{Phase diagram of Ho films with 10
monolayers.}\label{Fig.4}
\end{figure}

\begin{figure}[h]
\includegraphics[scale=0.27]{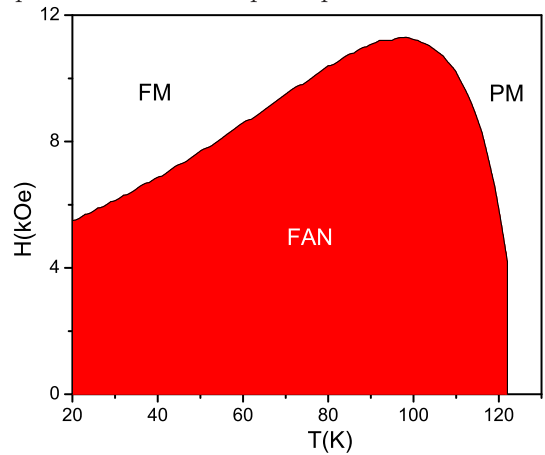}
\caption{Phase diagram of Ho films with 7 monolayers.}\label{Fig.5}
\end{figure}

Our results evince that in ultrathin Ho films ($N<10$ monolayers), the
helimagnetic phase is not stabilized. We can see in
Fig.(\ref{Fig.5}) that for ultrathin films the interval of magnetic
field, which leads to a phase transition, is smaller than the
interval of magnetic field in Ho bulk and thick Ho films. In this
case the specific heat measure may be not a good tool for the
identification of the magnetic phase transition. Also, from the
Figs. (4-a), (5), and (6), we can observe the appearing of the  helifan
phase, as we increase the number of monolayers. Also, as we have
commented above, this is a surface-induce effect. This effect is expected for
this type of magnetic thin films, as it was reported in other works \cite{Carrico,Mello}. With the
increasing of the number of monolayers ($N \gg 1$) this
helifan phase will disappear, recovering the $H$-$T$ diagram of Ho
in bulk, as in Fig. (\ref{phase1}).

\section{Conclusion}

In conclusion, we have studied the magnetic phases of very thin Ho
films in the temperature interval between 20K and 132K. The present
study shows the strong influence that the surface and thickness of a
thin film (See Figs. (4), (5), and (6)), associated with the existing
competition between the energies of exchange and magneto-crystalline
anisotropy, exerts on the magnetic order of these systems, when in the presence
of an external magnetic field and temperature. The slab
size, surface effects and magnetic field due to spin ordering impact
significantly the magnetic phase diagram. Also, the presence of an
external field gives rise to the magnetic phase Fan and the
spin-slip structures observed in Ho originates from the anisotropy
competition between magnetic hexagonal and energy exchange energies,
causing a significant change in the magnetic symmetry of the system
(Figs 1-6). Specifically, the helifan phase emerges due to the
presence of surfaces, and to a limit where the film thickness is
twice the period of the helix of the film and have a strong planar
anisotropy of the same order of magnitude of the exchange energy
(Fig. 4). The specific heat curves is presented, which consider both
magnetic and elastic (lattice) contributions. From Fig. (2) we can see
that all magnetic phase transitions, for strong enough magnetic fields, are marked by the specific heat.
Here we emphasize that the specific heat measurements in weak
magnetic fields do not mark all magnetic phase transitions, because
the field interval that destabilizes the magnetic order is low, so we consider the magnetic susceptibility measurements as more appropriate phase markers, for this interval of fields.
\begin{acknowledgments}
The authors acknowledge the financial support from the Brazilian
research agencies CNPq and FAPERN.
\end{acknowledgments}


\end{document}